\documentclass[%
 reprint,
 amsmath,amssymb,
 aps,
]{revtex4-2}

\usepackage{accents}
\usepackage{graphicx}
\usepackage{dcolumn}
\usepackage{bm}

\usepackage[utf8]{inputenc}
\usepackage[english]{babel}
\usepackage{amsmath}
\usepackage{amsthm}
\usepackage{amssymb}
\usepackage{makecell} 
\usepackage{graphicx}
\graphicspath{ {./images/} }
\usepackage[margin=1in]{geometry}
\usepackage[english]{babel}
\usepackage[utf8]{inputenc}
\usepackage{physics}
\usepackage{float}
\usepackage{setspace}
\usepackage{stackengine}
\usepackage[margin=1in]{geometry}
\usepackage{tikz}
\usetikzlibrary{positioning}
\usepackage{accents}
\usepackage{wrapfig}
\usepackage[export]{adjustbox}
\usepackage{xcolor}
\usepackage{soul}
\usepackage{yfonts}
\usepackage{array}
\usepackage{booktabs}

\newcommand{\PreserveBackslash}[1]{\let\temp=\\#1\let\\=\temp}
\newcolumntype{C}[1]{>{\PreserveBackslash\centering}p{#1}}
\newcolumntype{R}[1]{>{\PreserveBackslash\raggedleft}p{#1}}
\newcolumntype{L}[1]{>{\PreserveBackslash\raggedright}p{#1}}

\makeatletter
\newcommand*\bigcdot{\mathpalette\bigcdot@{.5}}
\newcommand*\bigcdot@[2]{\mathbin{\vcenter{\hbox{\scalebox{#2}{$\m@th#1\bullet$}}}}}
\makeatother

\newcolumntype{Q}[1]{>{\centering\arraybackslash}m{#1}}

\begin{document}

\preprint{APS/123-QED}

\title{Pseudo-Hermitian physics from dynamically coupled macrospins}

\author{Peter Connick}

\author{Shane P. Kelly}

\author{Yaroslav Tserkovnyak}

\affiliation{Department of Physics and Astronomy and Bhaumik Institute for Theoretical Physics, University of California, Los Angeles, California 90095, USA}

\date{\today}

\begin{abstract}

We consider two classical macrospins with dynamical (frequency-dependent) coupling, modeled by a generalized Landau-Lifshitz-Gilbert equation. We show that, in the absence of local damping, the resulting dynamics are pseudo-Hermitian.  When two precessional modes hybridize near a crossing, the spectral behavior takes the form either of an anticrossing or level attraction, with the latter formalized in terms of spontaneous $\mathcal{PT}$-symmetry breaking. Near equilibrium, mixing due to nondissipative interactions results in repulsion, while dissipative mixing results in attraction. In contrast, when the fluctuating degrees of freedom form a free-energy saddle point, we find that nondissipative interactions result in level attraction, while dissipative interactions produce level repulsion.  Accounting for the effects of local Gilbert damping, we examine the cases in which approximate $\mathcal{PT}$-symmetry breaking is still possible and determine the degree to which the qualitative spectral properties still persist.

\end{abstract}

\maketitle


\section{Introduction}

Spintronics is the fundamental study and engineering of spin degrees of freedom for communication, information and energy-storage purposes \cite{hirohata2020review, dieny2020opportunities, duine2011alternating, chumak2015magnon}.
In these applications, spin waves carry both information and energy, and are appealing due to the potentially low-energy dissipation and radio-frequency compatibility.
At low temperatures and short wavelengths, the quanta of spin waves, magnons, have also generated interest for their potential utility to quantum sensing and communication \cite{awschalom2013quantum, yuan2022quantum}. 

In the majority of applications, spin waves and magnons are linear excitations on top of an equilibrium ferromagnetic~\cite{harms2022antimagnonics, herring1951theory, dong2023collective} or antiferromagnetic~\cite{baltz2018antiferromagnetic, kim2014propulsion} order.
Close to equilibrium, the dispersion of spin waves is usually captured by static couplings, such as the exchange and magnetic-dipole interactions \cite{rezende2020fundamentals, herring1951theory}, while relaxation is often described by a viscous Gilbert damping within the Landau-Lifshitz-Gilbert (LLG) formalism \cite{LL}.  
In contrast, dynamical interactions, such as nonlocal Gilbert damping, are less frequently considered but can be relevant in both metallic systems \cite{tserkovnyak2005nonlocal, heinrich2003dynamic} and insulating systems; the latter being the study of recent experiments \cite{subedi2023magnon, yabinUnpublished}.

Furthermore, local Gilbert damping may be compensated by nonequilibrium spin torques using a variety of means~\cite{slonczewski1996current, mihai2010current, myers1999current, katine2000current}, giving rise to new regimes of collective dynamics governed by nonlocal coupling effects, both dissipative and nondissipative.
In these regimes, the linearized precession is generally nonconservative and dynamic instabilities are possible that may completely rearrange the magnetic order. We are interested in classifying the ensuing coupled dynamics in terms of the emergent non-Hermitian characteristics, rooted in Onsager reciprocities and collective dissipation. With weakly-coupled mode hybridization constituting a fundamental building block in understanding collective dynamics, we specifically focus on the fate of the spectral crossing points of two macrospins.

The level repulsion (anticrossing) can be formalized in terms of the underlying quasi-Hermiticity, while the level attraction can be formalized in terms of pseudo-Hermiticity and the associated spontaneous parity-time ($\mathcal{PT}$) symmetry breaking \cite{bender2019pt}. We will see how these distinct scenarios depend on the energetics and dissipative characteristics of the dynamical system. When constrained by Onsager reciprocity, the $\mathcal{PT}$-symmetry breaking occurs when the local frequencies are close and either 1) the dynamic couplings are nondissipative and the fixed point is at a free-energy saddle point or 2) when the dynamic couplings are dissipative and the fixed point is at a free-energy extremum (either minimum or maximum). Swapping the dissipative or energetic characteristics here would result in quasi-Hermiticity. We will conclude our discussion with perspectives for multispin dynamics and the associated neuromorphic computing.

In Sec.~II, we establish the physical model and construct pseudo-Hermitian perturbation theory, putting the phenomena of level attraction and repulsion in a formal context.
In Sec.~III, invoking Onsager reciprocity, we show that in the nondissipative case our model is generally pseudo-Hermitian (with quasi-Hermiticity being its special simpler case).
In Sec.~IV, we consider purely dissipative interactions and show that they too lead to pseudo-Hermitian dynamics, under certain assumptions.
In Sec.~V, we discuss some concrete examples and compute the relevant level splittings explicitly.  Finally, in Sec.~VI, we reintroduce finite local Gilbert damping and explore its impact on the preceeding results.

\section{Framework}

\begin{figure}[t]

\includegraphics[width=\columnwidth]{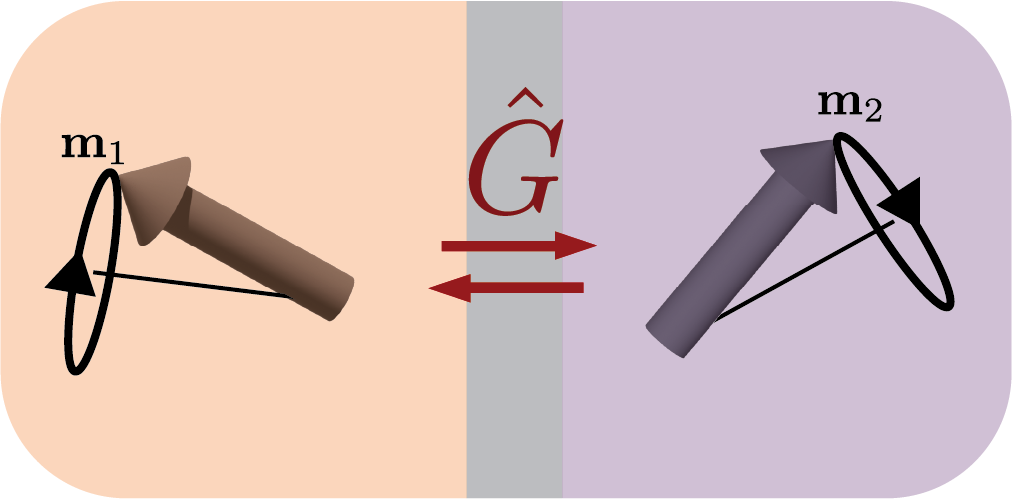}
\caption{Magnetic bilayer with spin precession modes -- a possible realization of our system. In this case, the dynamical interaction $\hat{G}$ is mediated through a metal spacer shown in gray.}
\label{fig:spincartoon}
\end{figure}

We consider two macrospins with fixed magnitudes $S_{1}$ and $S_2$ and a free energy that may in general account for external magnetic fields $\textbf{B}_\imath$, anisotropy $\hat{K}_\imath$, and exchange $J$:
\begin{equation}
\label{eqn:freeEnergy}
\begin{aligned}
    F= & -\textbf{B}_1 \cdot \textbf{S}_1-\textbf{B}_2 \cdot \textbf{S}_2\\
    &+ \frac{1}{2}\left( \textbf{S}_1 \cdot \hat{K}_1 \textbf{S}_1 +\textbf{S}_2 \cdot \hat{K}_2 \textbf{S}_2\right) + J \textbf{S}_1 \cdot  \textbf{S}_2. \\
\end{aligned}
\end{equation}
The low-frequency precessional dynamics are described by the Landau-Lifshitz-Gilbert equation:
\begin{equation}
\label{eqn:EoM}
(1+ \alpha_{\imath} \textbf{m}_{\imath} \times) \dot{\textbf{m}}_{\imath} + \textbf{B}^{\text{eff}}_{\imath} \times \textbf{m}_{\imath} = \hat{G}_{\imath}\left(\textbf{m}_1, \textbf{m}_2\right) \dot{\textbf{m}}_{\overline{\imath} },    
\end{equation}
where $\overline{\imath} = 2,1$ for $\imath = 1, 2$. $\alpha_{\imath}$ parametrizes local Gilbert damping at site $\imath$, $\textbf{B}^{\text{eff}}_{\imath}\equiv -\delta F / \delta \textbf{S}_{\imath}$, and $\mathbf{m}_\imath\equiv\textbf{S}_\imath/S_\imath$.

The term $\hat{G}_{\imath}\left(\textbf{m}_1, \textbf{m}_2\right)\dot{\textbf{m}}_{\overline{\imath} }$ captures a dynamic interaction between the spins. 
Notably, due to its proportionality to $\dot{\textbf{m}}_{\overline{\imath}}$,  it does not affect stationary configurations of the system, distinguishing it from static couplings such as exchange. 
Such torques have been studied in metallic systems \cite{tserkovnyak2005nonlocal, heinrich2003dynamic}, and recent experimental work has suggested that they may be relevant in insulating systems as well \cite{subedi2023magnon, yabinUnpublished}. 

As illustrative examples, we consider two leading-order isotropic dynamic couplings:
\begin{subequations}
\label{eqn:exampleDef}
  \begin{align}
    \hat{G}_{\imath}^- \dot{\textbf{m}}_{\overline{\imath} } &= \frac{S}{S_{\imath}} \alpha' \textbf{m}_{\imath} \times \dot{\textbf{m}}_{\overline{\imath} } \label{eqn:exampleDefa},\\
 \hat{G}_{\imath}^+ \dot{\textbf{m}}_{\overline{\imath} } &= \frac{S}{S_{\imath}}\textbf{m}_{\imath} \times \left[(\eta_{\imath} \textbf{m}_{\imath} + \eta_{\overline{\imath} } \textbf{m}_{\overline{\imath} }) \times \dot{\textbf{m}}_{\overline{\imath} }  \right], \label{eqn:exampleDefb}
  \end{align}
\end{subequations}
where $S = \sqrt{S_1 S_2} $ is the geometric average of the spin magnitudes. 
These interactions are constructed by identifying the lowest-order in $\textbf{m}_i$ couplings that preserve the magnitude of both spins and respect both SO(3) symmetry and Onsager reciprocity. (See Appendix A for the latter). 
The superscripts $\pm$ indicate whether the interaction preserves or breaks the time-reversal symmetry in Eq.~(\ref{eqn:EoM}): under time-reversal, the magnetization reverses $\textbf{m}_\imath\rightarrow -\textbf{m}_\imath$ and dynamical interactions in Equations~(\ref{eqn:exampleDefa}) and (\ref{eqn:exampleDefb}) satisfy $\hat{G}^{\pm}_i(-\textbf{m}_1, -\textbf{m}_2) = \pm  \hat{G}^{\pm}_i(\textbf{m}_1, \textbf{m}_2)$.
Note that Eq.~(\ref{eqn:exampleDef}) does not depend on an applied magnetic field; this ensures that the dynamical coupling respects Onsager reciprocity for any free energy of the magnets. This can be justified by assuming that the mediator of the dynamical coupling is weakly coupled to the external magnetic field~(relative to the coupling with the macrospins)~\cite{tserkovnyak2005nonlocal}.

This work is primarily concerned with the linear dynamics near the fixed points $\textbf{m}_{0,\imath}$ of Eq.~(\ref{eqn:EoM}), which are identified by $\delta F / \delta \textbf{m}_{\imath}= 0$.
We take $\textbf{m}_{\imath}(t) = \textbf{m}_{0,\imath} + \delta \textbf{m}_{\imath}(t)$, where $\delta \textbf{m}_{\imath}(t)$ is understood to be small, and, for ease of notation, combine these fluctuations into a single four-component column vector denoted by the gothic symbol  $\delta \textswab{m} \equiv (\delta \text{m}_{x,1}, \delta \text{m}_{y,1}, \delta \text{m}_{x,2}, \delta \text{m}_{y,2} )^{\intercal}$.
To linear order, the dynamics of Eq.~(\ref{eqn:EoM}) take the form of a Schr\"{o}dinger equation with non-Hermitian Hamiltonian $\hat{h}$:
\begin{equation}
    \label{eqn:schrodinger}
     i\delta \dot{\textswab{m}} = \hat{h} \delta \textswab{m}.
\end{equation}
This Hamiltonian is said to be \textit{pseudo-Hermitian} if it satisfies
\begin{equation}
    \label{eqn:pseudoHermitian}
    \hat{\Theta} \hat{h} \hat{\Theta}^{-1} = \hat{h}^{\dagger},
\end{equation}
for some similarity transformation $\hat{\Theta}$, which we call the \textit{pseudometric.} For a brief overview of the fundamentals of pseudo-Hermiticity, see Appendix B.
In particular, $\hat{\Theta}$ may generally be chosen to be Hermitian and thus have real eigenvalues, which we will assume henceforth. 
Equation~(\ref{eqn:pseudoHermitian}) is equivalent to the existence of an antilinear symmetry known as a $\mathcal{PT}$ symmetry, which guarantees that the complex-valued eigenvalues of $\hat{h}$ come in conjugate pairs.
Below, we discuss the perturbation theory and general spectral properties of pseudo-Hermitian systems in the vicinity of a spectral crossing.

\subsection*{Pseudo-Hermitian perturbation theory}

\renewcommand{\arraystretch}{1.5}
\begin{table*}
\centering
\begin{tabular}{| C{4cm} || C{3cm} | C{3cm} | C{3cm}| } 
 \hline
 & Hermitian  & Quasi-Hermitian & Pseudo-Hermitian \\ [0.5ex] 
 \hline\hline
 Symmetry & $ \hat{H} = \hat{H}^{\dagger} $ & $\hat{\Theta}\hat{H}\hat{\Theta}^{-1} = \hat{H}^\dagger$ & $\hat{\Theta}\hat{H}\hat{\Theta}^{-1} = \hat{H}^\dagger$ \\ 
 \hline
 Conserved ``probability" &  $\langle \psi | \psi \rangle>0$ &  $\langle \psi | \hat{\Theta} | \psi \rangle > 0$ & $ \langle \psi | \hat{\Theta} | \psi \rangle $\\
 \hline
 Spectrum & real & real & conjugate pairs \\
 \hline
\end{tabular}
\caption{Notable classes of Hamiltonians and their properties. We have assumed $\psi\neq0$, and $\hat{\Theta}$ is chosen to be Hermitian and, for quasi-Hermitian Hamiltonians, strictly positive. Hermitian matrices are a strict subset of quasi-Hermitian matrices, which are a strict subset of pseudo-Hermitian. Each class possesses a conserved quantity, which, in the Hermitian and quasi-Hermitian cases, is strictly positive and thus may be interpreted as a probability.}
\label{table:1}
\end{table*} 
 
We address the general problem in which two eigenmodes hybridize near a level crossing in the context of  pseudo-Hermitian dynamics. Projecting onto the subspace of the crossing modes, we formalize the concepts of level repulsion and attraction, either of which may occur depending on the pseudometric structure. The problem is formulated in terms of a perturbation theory for a Hamiltonian of the form
\begin{equation}
    \label{Hperturbed}
    \hat{h} = \hat{h}_0 + \hat{h}'.
\end{equation}
Here, the unperturbed part $\hat{h}_0$ is assumed to be pseudo-Hermitian with pseudometric $\hat{\Theta}$ and to have real-valued frequencies, as in our physical cases of interest below. We remark that while the pseudometric of $\hat{h}_0$ is generally not unique, the form of the perturbation $\hat{h}'$ may dictate its natural choice, as we will see in our examples. The perturbation $\hat{h}'$, which is responsible for mixing eigenstates of $\hat{h}_0$, is generally not required to be pseudo-Hermitian in its own right.

\begin{figure}[b]
\centering
\includegraphics[width=\columnwidth]{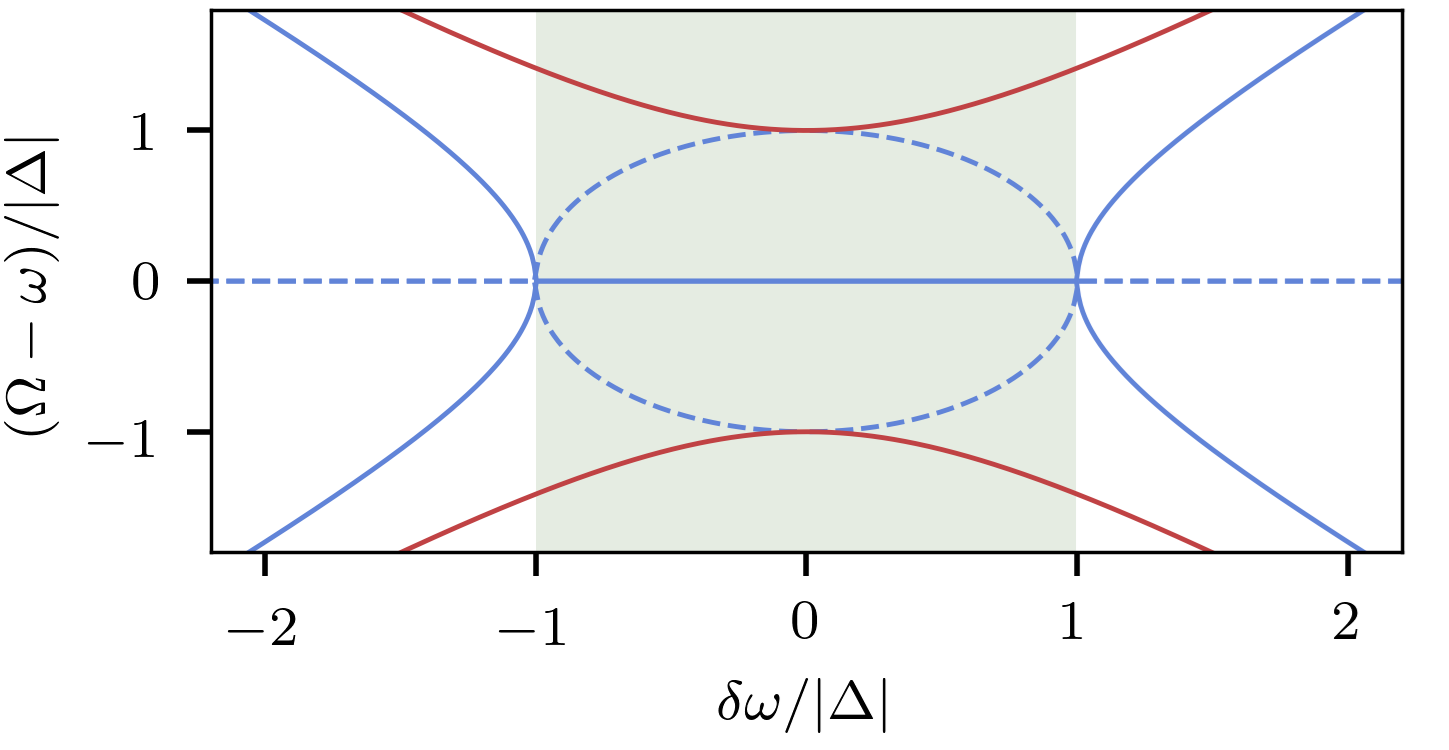}
\caption{Frequency splitting for dynamically coupled modes as described by Eq.~(\ref{eqn:eigenvalues}).  The red curve corresponds to the case where $\theta_1 \theta_2 = 1$.  The blue curves correspond to the case where $\theta_1 \theta_2  = -1$, with solid lines indicating the real part and dashed lines indicating the imaginary part. Spontaneous $\mathcal{PT}$ symmetry breaking occurs when $\theta_1 \theta_2 = -1$ in the green shaded region, beginning at the exceptional points where $\delta \omega = \pm \Delta$.}
\label{fig:ptbreak}
\end{figure}

As discussed in Ref.~\cite{mostafazadeh2008pseudo}, for any pseudo-Hermitian Hamiltonian with real eigenvalues and away from any exceptional points, one can introduce the Riesz basis of right and left eigenvectors, denoted respectively by $\{|\psi_n\rangle \}$ and $\{|\phi_n \rangle \}$ and satisfying 
\begin{equation} 
\label{eqn:orthogonal}
\langle\phi_n | \psi_m\rangle =\delta_{n m}.
\end{equation}
Here, $\langle\phi_n |$ is the usual conjugate transpose of $|\phi_n \rangle $, and the index runs over $n =1,...,N$, with $N$ being the dimension of the vector space.
The projector onto $|\psi_n \rangle $ is given by the pseudo-Hermitian operator $| \psi_n \rangle \langle \phi_n|$. 
The generalization of this basis for pseudo-Hermitian Hamiltonians with complex eigenvalues is discussed in Appendix B.  

We develop perturbative treatment relative to the basis of the $\hat{h}_0$ eigenvectors, which satisfy
\begin{equation}
    \label{eqn:LReigen}
    \langle \phi_n | \hat{h}_0 = \langle \phi_n | \omega_n  , ~~~ \hat{h}_0 | \psi_n \rangle = \omega_n | \psi_n \rangle,
\end{equation}
where $\omega_n$ are real.
As discussed in Appendix B, the left and right eigenvectors are related by $\hat{\Theta}|\psi_n\rangle =  \theta_n |\phi_n\rangle$, where the expectation value $\theta_n = \langle \psi_n | \hat{\Theta}| \psi_n \rangle$ is guaranteed to be nonzero for eigenmodes with real frequencies. This suggests that we may choose normalization of $|\psi_n \rangle$ such that $\theta_n = \pm 1$, which we will do henceforth.  Furthermore, we define $\theta_n $ to be the sign of the eigenmode $|\psi_n\rangle$.  
 
We now consider a level crossing of the unperturbed eigenmodes $|\psi_1 \rangle $  and $|\psi_2 \rangle $, where the corresponding eigenfrequencies satisfy 
\begin{equation}
    \label{eqn:crossingcondition}
    \left| \omega_1 - \omega_2 \right| \lesssim | \langle \phi_1 | \hat{h} | \psi_2 \rangle |,
\end{equation}
while all other frequencies are far separated from $\omega_1 $ and $\omega_2$ compared to their respective couplings.  
Near the crossing, the dynamics are dominated by the projected Hamiltonian
\begin{equation}
\label{eqn:ProjectedHamiltonian}
\hat{h}_{2 \times 2}=\left(\begin{array}{ll}
\langle \phi_1 |\hat{h}| \psi_1\rangle & \langle \phi_1|\hat{h}| \psi_2\rangle \\
\langle\phi_2|\hat{h}| \psi_1\rangle &  \langle \phi_2|\hat{h}| \psi_2 \rangle
\end{array}\right).
\end{equation}
The diagonal elements $\langle \phi_{\imath} |\hat{h}| \psi_{\imath}\rangle$ are equal to $\omega_{\imath}$, in the absence of the perturbation $\hat{h}'$. 

We now specialize to the case that the projected perturbation $\hat{h}'_{2\times 2} $ is also pseudo-Hermitian with a pseudometric $\hat{\Theta} $ that is shared with $\hat{h}_0$.  For most of the examples that we will consider, it will turn out that the entire perturbation $\hat{h}'$ is pseudo-Hermitian, but for our purposes this is only necessary for the projected part. 
In this case, we make the substitution for the off-diagonal elements $\langle \phi_{\imath} |\hat{h}| \psi_{\overline{\imath}}\rangle = \theta_{\overline{\imath}} \langle \phi_{\imath} |\hat{h}\hat{\Theta}^{-1}| \phi_{\overline{\imath}}\rangle$ and note that $\langle \phi_{\imath} |\hat{h}\hat{\Theta}^{-1}| \phi_{\overline{\imath}}\rangle = \langle \phi_{\overline{\imath}} |\hat{h}\hat{\Theta}^{-1}| \phi_{\imath}\rangle^*$.
Calculation of the eigenfrequencies is straightforward, resulting in 
\begin{equation}
    \label{eqn:eigenvalues}
    \Omega = \omega \pm \sqrt{\delta \omega^2 + \theta_1 \theta_2 | \Delta |^2},
\end{equation}
where $\omega =  (\langle \phi_1|\hat{h}| \psi_1\rangle + \langle \phi_2|\hat{h} | \psi_2\rangle)/2$,  $\delta \omega = (\langle \phi_1|\hat{h} | \psi_1\rangle - \langle \phi_2|\hat{h} | \psi_2\rangle)/2$, and $\Delta = \langle \phi_1 |\hat{h}\hat{\Theta}^{-1}| \phi_2\rangle $.  
The eigenfrequencies given by Eq.~(\ref{eqn:eigenvalues}) are plotted in Fig.~(\ref{fig:ptbreak}).
For modes of the same sign, the argument of the square root function in Eq.~(\ref{eqn:eigenvalues}) is positive, and hybridization takes the form of an anticrossing with gap size $| \Delta |$.
Conversely, for modes of opposite sign, the argument of the square root becomes negative and the real frequencies degenerate when $\delta \omega^2 < | \Delta |^2 $. 
From the perspective of $\mathcal{PT}$ symmetry, this level attraction indicates a region where the eigenstates of the Hamiltonian cease to be eigenstates of the $\mathcal{PT}$ operator, in which case it is said that the symmetry has been ``spontaneously broken" \cite{bender2019pt}.

\section{Nondissipative coupling}

We establish pseudo-Hermiticity for the dynamics associated with dissipationless torques, which, according to the Onsager-reciprocal relations, are identifiable as those which preserve time reversal symmetry. 
We find that for fluctuations near free-energy extrema (and, in particular, near equilibrium), hybridization generally occurs in the form of anticrossings. 
Alternatively, hybridizing fluctuations about free-energy saddle points may result in level attraction.
 
For linear fluctuations near a fixed point, the Onsager matrix of kinetic coefficients $\hat{\Gamma}$ is defined as the real-valued matrix relating the velocities to the generalized forces:
\begin{equation}
\label{eqn:Gammadef}
    \delta \dot{\textswab{m}} \equiv \hat{\Gamma} \, \textswab{f}, ~~~ \textswab{f} \equiv -\hat{\beta}\, \delta \textswab{m}.
\end{equation}
Here, the real-valued symmetric matrix $\hat{\beta} $ is the free-energy curvature defined by 
\begin{equation}
    \label{eqn:betadefF}
    F = F_0+ \frac{1}{2}\delta \textswab{m}^{\intercal} \  \hat{\beta} \ \delta \textswab{m} + O(\delta \textswab{m}^3),
\end{equation}
with $F_0$ representing the free energy at the fixed point. 
As discussed in Ref.~\cite{LL}, the Onsager reciprocity, which relies on the microscopic time-reversal symmetry, dictates that
\begin{equation}
\label{eqn:OnsagerReciprocity}
     \hat{\Gamma} (-\textbf{m}_{\imath}, -\textbf{B}_{\imath}) = \hat{\Gamma}(\textbf{m}_{\imath}, \textbf{B}_{\imath})^{\intercal}.
\end{equation}
Strictly speaking, these relations are only required to hold near equilibrium.  
However, dynamic interactions that are essentially independent of the free energy, such as Eqs.~(\ref{eqn:exampleDefa}) and (\ref{eqn:exampleDefb}), should satisfy Eq.~(\ref{eqn:OnsagerReciprocity}) for expansions about any orientation. 

Comparing Eqs.~(\ref{eqn:Gammadef}) and (\ref{eqn:schrodinger}), we relate the Onsager matrix to the effective Hamiltonian as
\begin{equation}
\label{eqn:hEqualsGamma}
    \hat{h} =- i \hat{\Gamma} \hat{\beta}.
\end{equation}
In order to identify the parts of $\hat{\Gamma}$ responsible for dissipation, it is useful to define the (anti)symmetric components $\hat{\Gamma}_{\pm} \equiv (\hat{\Gamma} \pm \hat{\Gamma}^{\intercal})/2 $, which according to Eq.~({\ref{eqn:OnsagerReciprocity}}) are, respectively, even/odd under time reversal. 
Energy loss may then be calculated as 
\begin{equation}
\label{eqn:EnergyDissipation}
    -\dot{F}\equiv \textswab{f}^{\intercal}\delta \dot{\textswab{m}}  = \textswab{f}^{\intercal}  \hat{\Gamma}_+ \textswab{f}.
\end{equation}
Thus, dissipation is determined by the even part, $\hat{\Gamma}_+$, which breaks time reversal symmetry in Eq.~(\ref{eqn:Gammadef}).

In the case without dissipation, we have $\hat{\Gamma} = \hat{\Gamma}_-$, and, consequently,
\begin{equation}
    \label{eqn:hEqualsGammaMinus}
    \hat{h} =- i \hat{\Gamma}_- \hat{\beta}.
\end{equation}
This Hamiltonian exhibits pseudo-Hermiticity (\ref{eqn:pseudoHermitian}) with pseudometric given by $\hat{\Theta} = \hat{\beta}$ up to an overall sign. 
A mode's sign is then precisely the sign of its $\hat{\beta}$ expectation value.
For fluctuations near free-energy extrema, $\hat{\beta}$ can only have expectation values of one sign and thus hybridization can only occur in the form of anticrossings; in this case, the dynamics are said to be quasi-Hermitian.
Alternatively, when the fluctuating degrees of freedom form a free-energy saddle point, modes of opposite sign are allowed to mix; thus $\mathcal{PT}$-symmetry breaking is possible sufficiently close to a crossing. 
These results are summarized in the second column of Table \ref{table:hybrid}. 

\section{Purely dissipative coupling}

\renewcommand{\arraystretch}{1.5}
\begin{table*}
\centering
\begin{tabular}{c}  
\includegraphics[width=0.85\linewidth]{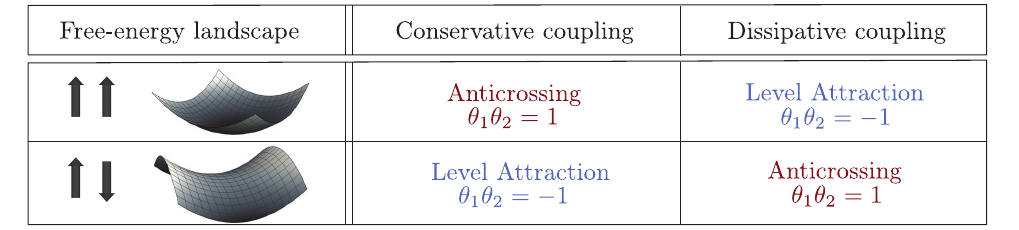}
\end{tabular}
\caption{Spectral behavior of spins with parallel magnetic fields hybridizing near a crossing. The left-most column specifies the relative orientation and free energy landscape, with parallel spin directions indicating a free energy extremum and antiparallel indicating a saddle point.  $\theta_{\imath}$ represents the sign of the ${\imath}$th mode according to the pseudometric.}
\label{table:hybrid}
\end{table*}

Purely dissipative interactions also exhibit pseudo-Hermitian physics in the absence of local damping and static couplings, although identifying the pseudometric requires a more detailed understanding of the Hamiltonian structure. 
The ultimate consequence is that, for the dynamics of two spins, the dependence on the free-energy curvature is reversed compared to the nondissipative case. 
Hybridization near equilibrium now results in $\mathcal{PT}$-symmetry breaking.  
Alternatively, anticrossings become possible when the relevant fluctuations form a free-energy saddle point.

We consider the model given by Eqs.~(\ref{eqn:freeEnergy})~and~(\ref{eqn:EoM}), but without any static coupling $(J=0)$.
We apply the linearization procedure discussed in Section II, after which the LLG equation takes the form 
\begin{equation}
    \label{eqn:EoMLinearized}
    (1 - \hat{A}) \delta \dot{\textswab{m}} + i\hat{h}_0 \delta \textswab{m} = \hat{G}\delta \dot{\textswab{m}},
\end{equation}
with the $4 \times 4$ matrices $\hat{h}_0$, $\hat{A}$, and $\hat{G}$ representing the linearizations of the effective magnetic field, local damping, and dynamic interaction, respectively. 
By comparing with their nonlinear counterparts in Eq.~(\ref{eqn:EoM}), it follows that $\hat{h}_0$ and $\hat{A}$ are odd under time reversal.

Equation~(\ref{eqn:EoMLinearized}) can be recast as a Schr\"{o}dinger equation with a non-Hermitian Hamiltonian by solving for $\delta \dot{\textswab{m}}$:
\begin{equation}
    \label{eqn:hFromEoM}
 \hat{h} = (1 - \hat{A} - \hat{G})^{-1} \hat{h}_0.
\end{equation}
For the remainder of this section, we formally consider the case without local damping ($\hat{A} = 0$), while emphasizing that this is physically relevant only when the magnets have been subjected to external antidamping torques~\cite{slonczewski1996current, mihai2010current, myers1999current, katine2000current}. We will restore local damping in Sec.~\ref{GD}.

As we see below, the Hamiltonian without local damping and with purely dissipative $\hat{G}$ satisfies
\begin{equation}
    \label{eqn:pseudoHermitianDissipative}
    \hat{Z}\hat{\beta} \hat{h} (\hat{Z}\hat{\beta})^{-1} = \hat{h}^{\dagger},
\end{equation}
where we define the operator $\hat{Z}$ by $\hat{Z} \delta \textswab{m} =  (\delta \text{m}_{x,1}, \delta \text{m}_{y,1}, -\delta \text{m}_{x,2}, -\delta \text{m}_{y,2} )^{\intercal}$.
This estbalishes pseudo-Hermitian dynamics, similar to the nondissipative case, but now with $\hat{\Theta} = \hat{Z}\hat{\beta}$.

To prove Eq.~(\ref{eqn:pseudoHermitianDissipative}), we identify the Onsager matrix by comparing Eqs.  (\ref{eqn:Gammadef}) and (\ref{eqn:hFromEoM}), which results in
\begin{equation}
    \label{eqn:GammaFromEoM}
    \hat{\Gamma} = (1-\hat{G})^{-1} \hat{\Gamma}_0.
\end{equation}
Here, $\hat{\Gamma}_0 = i\hat{h}_0 \hat{\beta}^{-1}$ is the Onsager matrix in the absence of all dynamical effects; as such, it is real-valued, antisymmetric, and odd under time reversal.
According to Onsager reciprocity, $\hat{\Gamma}^{\intercal}$ (and  by extension $\hat{\Gamma}_{\pm}$) may be obtained by time reversing Eq.~(\ref{eqn:GammaFromEoM}), i.e., calculating it for the reversed magnetic fields and magnetizations.  
We now take the dynamic interaction to be purely dissipative, by which we mean that it satisfies
\begin{equation}
    \label{eqn:pureDissipativeCondition}
    \hat{G}(-\textbf{m}_1, -\textbf{m}_2) = -\hat{G}(\textbf{m}_1, \textbf{m}_2),
\end{equation}
thus breaking time-reversal symmetry in Eq.~(\ref{eqn:EoM}).  
By applying time reversal to Eq.~(\ref{eqn:GammaFromEoM}) and utilizing Onsager reciprocity, it follows that the symmetric and antisymmetric components of $\hat{\Gamma}$ are given by
\begin{equation}
\label{eqn:DissipativeGammaSplit}
    \hat{\Gamma}_+ = \frac{\hat{G}}{1 - \hat{G}^2} \hat{\Gamma}_0, ~~~ \hat{\Gamma}_- = \frac{1}{1 - \hat{G}^2} \hat{\Gamma}_0.
\end{equation}
Thus, the full Onsager matrix satisfies
\begin{equation}
    \label{eqn:GammaPseudoHermitian}
    \hat{Z} (i\hat{\Gamma}) \hat{Z} = (i \hat{\Gamma})^{\dagger}.
\end{equation}
$\hat{Z}$ has the effect of changing the signs of nonlocal operators, so that $\hat{Z}\hat{\Gamma}_0 \hat{Z} =  \hat{\Gamma}_0$ and  $ \hat{Z}\hat{G}\hat{Z} = -\hat{G} $.
Due to the absence of free-energy couplings, $\hat{\beta}$ is a purely local operator, and thus we also have $\hat{Z} \hat{\beta} \hat{Z} = \hat{\beta}$.
This, in conjunction with Eq.~(\ref{eqn:GammaPseudoHermitian}), implies Eq.~(\ref{eqn:pseudoHermitianDissipative}).

With pseudo-Hermiticity established, the eigenfrequencies near a crossing are again given by Eq.~(\ref{eqn:eigenvalues}).  
The difference in the dissipative case lies in the physical meaning of the signs $\theta_{\imath}$, which are no longer equal to the signs of the free-energy curvature.  
Instead, because of the extra factor of $\hat{Z}$ in the pseudometric, the sign of the second spin's fluctuations is negative (positive) when expanded about a free-energy minimum (maximum), while the reverse is true for the first spin.
Thus, if the full system is near a free-energy extremum, we have $\theta_1 = -\theta_2$, leading to $\mathcal{PT}$-symmetry breaking when the precession modes hybridize.  
Alternatively, if one spin is expanded about a free energy minimum and the other is expanded about a maximum, then we have $\theta_1 = \theta_2$ and only anticrossings may occur.  
These results are summarized in the third column of Table \ref{table:hybrid}.

\section{Example: colinear case}
As a simple example, we consider the case of colinear magnetic fields oriented in the $z$ direction with no exchange or anisotropy ($K = J = 0$). We start by setting $\alpha_{\imath} = 0$, postponing the case with $\alpha_{\imath} \neq 0$ until Sec.~VI.  The stationary points of Eq.~(\ref{eqn:EoM}) are now given by orientations $\textbf{m}_{0,\imath} = \zeta_{\imath} \mathbf{z}$, where $\zeta_{\imath} = \mp 1$ indicates whether the $\imath$th spin is (anti)aligned with the magnetic field.
From an energetic perspective, opposite values for the $\zeta_{\imath}$ variables indicate a saddle point, while equal  values indicate an extremum for the full system (see Table \ref{table:hybrid}).
To linear order in the interaction $\hat{G}$, the Hamiltonian $\hat{h}$ in Eq.~(\ref{eqn:hFromEoM}) takes the form 
\begin{equation}
    \label{eqn:hLinearG}
    \hat{h} \approx \hat{h}_0 + \hat{G}\hat{h}_0.
\end{equation}
Explicit forms for $\hat{h}_0$, $\hat{G}$, and the free-energy curvature $\hat{\beta}$ are given in Appendix C.  

\subsection*{Nondissipative interaction}

Here, we compute the unperturbed right and left eigenvectors and the level splitting for the nondissipative interaction given by Eq.~(\ref{eqn:exampleDefb}).  
In the absence of disipation, we immediately identify the pseudometric of the full Hamiltonian as $\hat{\Theta} = \hat{\beta}$, up to an overall sign.  
The $U(1)$ symmetry in the colinear case implies that the eigenstates of $\hat{h}_0$ will undergo circular precession and thus are given by $|\psi_{1,\pm} \rangle = (1,\pm i, 0,0)^{\intercal}/\sqrt{2 S_1 \omega_1}$ and $|\psi_{2,\pm} \rangle = (0,0,1,\pm i)^{\intercal}/ \sqrt{2 S_2 \omega_2}$.
For a general fluctuation $\delta \textswab{m}$, the $\hat{\Theta}$ expectation value may be computed as 
\begin{equation}
    \label{eqn:normOfDeltam}
    \delta \textswab{m}^{\intercal} \hat{\Theta}\delta \textswab{m}  = \zeta_1 B_1 S_1 \delta \textbf{m}_1^2 + \zeta_2 B_2 S_2 \delta \textbf{m}_2^2,
\end{equation}
from which we identify that the sign of $|\psi_{\imath,\pm} \rangle$ is $\theta_{\imath, \pm} = \zeta_{\imath}$.  

The left eigenstates in general may be computed using $|\phi_{\imath, \pm} \rangle = \theta_{\imath} \hat{\Theta} |\psi_{\imath,\pm}\rangle$, which, in this case, are proportional to the right eigenstates:
\begin{equation}
    \label{eqn:LeftFromRight}
    |\phi_{\imath,\pm} \rangle = S_{\imath}\omega_{\imath}|\psi_{\imath,\pm} \rangle.
\end{equation}
Recalling standard quantum mechanics, the left and right eigenstates may generally be chosen to be the same as long as the unperturbed Hamiltonian is Hermitian.

Because of the $U(1)$ symmetry, the left- and right-handed eigenstates decouple, such that the hybridization can only occur between the $|\psi_{\imath,+} \rangle$ or $|\psi_{\imath,-} \rangle$ modes.  
The level splitting is then given by
\begin{equation}
\label{eqn:DeltaNonDissipative}
    |\Delta| =   |\langle \phi_{1,\pm}|\hat{G}\hat{h}_0|\psi_{2,\pm}\rangle| = |\eta_1 \zeta_1 + \eta_2 \zeta_2 |\sqrt{\omega_1 \omega_2}.
\end{equation}
Notably, $\theta_{1,\pm}\theta_{2,\pm} = \zeta_1 \zeta_2$, implying that the level splitting is real in the aligned case and imaginary in the antialigned case (see Table \ref{table:hybrid}).

Note that the quantity defined by Eq.~(\ref{eqn:normOfDeltam}) is conserved for any pseudo-Hermitian Hamiltonian with pseudometric $\hat{\Theta}$.
We recognize this as the free energy expanded in the neighborhood of the fixed point. Dynamic instabilities are allowed by Eq.~(\ref{eqn:normOfDeltam}) for the antialigned configuration, $\zeta_1 = - \zeta_2$.  
In such a case, the free energy is expanded about a saddle point, where $\delta \textbf{m}_\imath^2$ may increase while still conserving energy.

\subsection*{Dissipative interaction}

The dissipative interaction $\hat{G}_{\imath}^{-}$ [Eq.~\eqref{eqn:exampleDefa}] also results in anticrossings and $\mathcal{PT}$-symmetry breaking, but now with the dependence on the spin orientations reversed. 
The former now occur in the antialigned case, while the latter occur when the spins are aligned.

Due to the dissipative nature of the interaction, the pseudometric of the full Hamiltonian is now expected to be $\hat{\Theta} = \hat{Z}\hat{\beta}$.  
The left and right eigenmodes remain the same as in the nondissipative case; however, due to the factor of $\hat{Z}$ in the pseudometric, the $\hat{\Theta}$ expectation value is now given by
\begin{equation}
    \label{eqn:normOfDeltamdis}
    \delta \textswab{m}^{\intercal} \hat{\Theta}\delta \textswab{m} = \zeta_1 B_1 S_1 \delta \textbf{m}_1^2 - \zeta_2 B_2 S_2 \delta \textbf{m}_2^2,
\end{equation}
and the signs of the modes become $\theta_{\imath,\pm} = (-1)^{\imath-1}\zeta_{\imath}$.
The calculation for the level splitting is also similar, resulting in 
\begin{equation}
\label{eqn:DeltaDissipative}
    |\Delta| =   |\langle\phi_{1,\pm}|\hat{G}\hat{h}_0|\psi_{2,\pm}\rangle| = \alpha'\sqrt{\omega_1 \omega_2}.
\end{equation}
In this case, we have $\theta_{1,\pm}\theta_{2,\pm} = -\zeta_1 \zeta_2$, and thus the level splitting is real in the antialigned case and imaginary in the aligned case.

Once again, it is useful to discuss the conserved quantity given by Eq.~(\ref{eqn:normOfDeltamdis}), which we recognize as the difference in energies of the two spins.
In the aligned case where $\zeta_1 = \zeta_2$, the fluctuations of each spin, $\delta \textbf{m}_{\imath}^2$, can increase while keeping their difference fixed.
Thus, exponentially growing modes are allowed near a free-energy extremum but not near a saddle point. 

\section{Gilbert Damping}
\label{GD}

We finally consider a physically realistic dissipatively coupled system evolving under its natural dynamics, where local Gilbert damping must be included to ensure stability of the free-energy minimum. While this will generally compromise the pseudo-Hermiticity, the local damping can, under certain conditions, merely result in an overall exponential decay envelope. The pseudo-Hermitian features can survive relative to that envelope.

As in the Section IV, we consider spins with a purely dissipative dynamical coupling and no static coupling.
However, we now include local Gilbert damping $\alpha_{\imath} \gtrsim O(\alpha')$ where $\alpha'$ is the dimensionless parameter defining the characteristic magnitude of the dissipative interaction $\hat{G}$, as in Eq.~\eqref{eqn:exampleDefa}.
Similar to IV, we assume that the coupling between the spins is purely dissipative and thus there is no coupling through the free-energy.
Expanding Eq.~(\ref{eqn:hFromEoM}) to linear order in the dynamic torques yields the approximate Hamiltonian 
\begin{equation}
    \label{eqn:hLinearWithAG}
    \hat{h} \approx (1 + \hat{G})\hat{h}_0 + \hat{A}\hat{h}_0.
\end{equation}
The first term satisfies the pseudo-Hermitian condition (\ref{eqn:pseudoHermitianDissipative}), while the second term represents the leading-order effects of local Gilbert damping. 

\begin{figure}

\includegraphics[scale = 1]{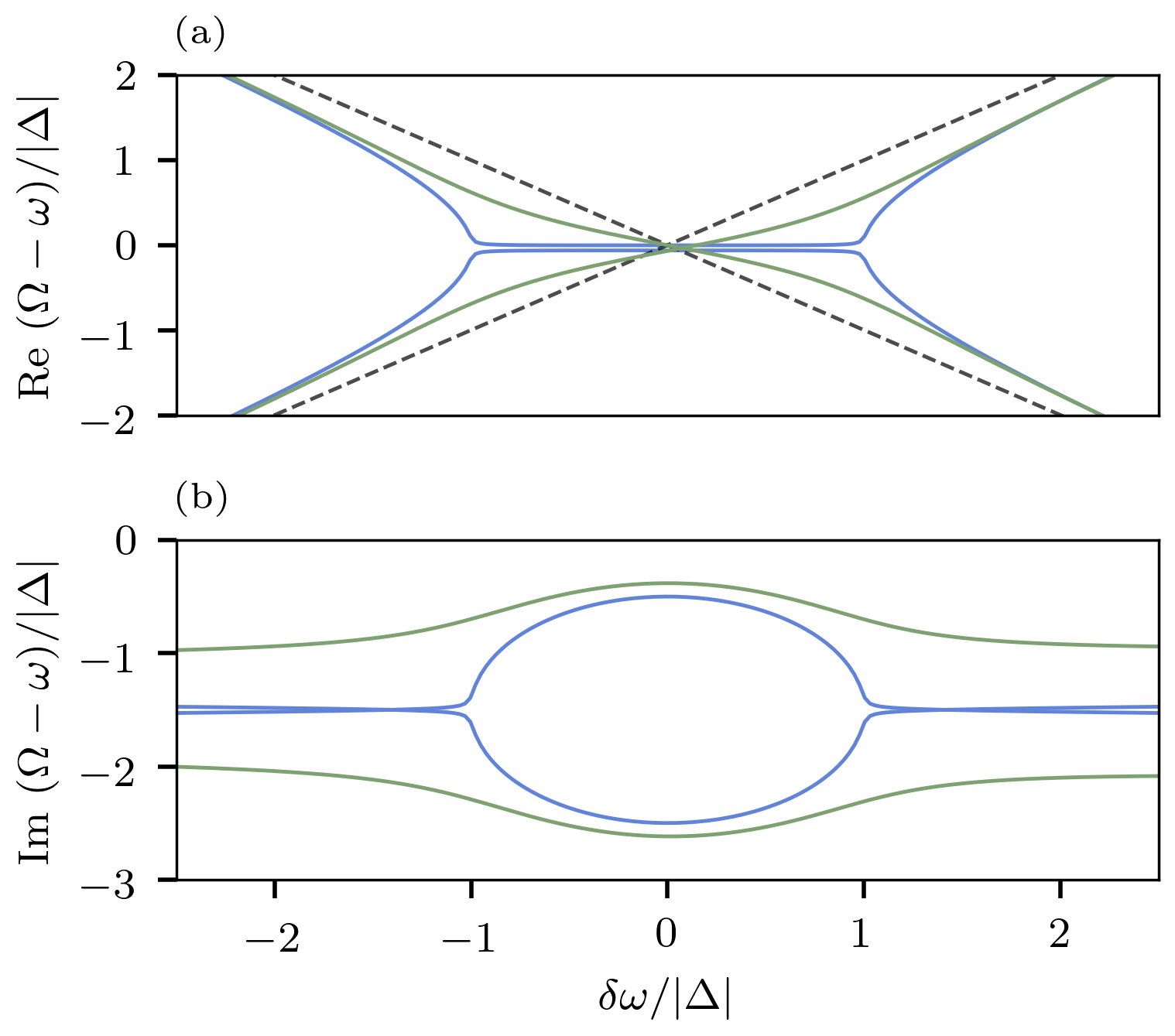}
\caption{Frequency splitting of spins with a dissipative interaction and local Gilbert damping.  Magnetic fields are chosen to be colinear, and the dissipative interaction is given by Eq.~(\ref{eqn:exampleDefa}), for two equal macrospins.  The panels (a) and (b) show the real and imaginary frequency components, respectively.  The case of symmetric Gilbert damping, with $\alpha_1 = \alpha_2 = 1.5\alpha'$ = 0.015, is shown in blue, while antisymmetric, with $\alpha_1 = 2 \alpha_2 = 2\alpha' = 0.02$, is shown in green.  The frequencies in the absence of all dynamic torques are shown in black (dashed), relative to which both the symmetric and asymmetric cases experience qualitative level attraction.}
\label{fig:gilPTbreak}
\end{figure}

We consider the regime of two-mode hybridization as defined in Eq.~(\ref{eqn:crossingcondition}), where the dynamics are well described by the projected Hamiltonian (\ref{eqn:ProjectedHamiltonian}).
Because the overall damping must result in imaginary frequencies of only one sign, the full dynamics cannot be $\mathcal{PT}$ symmetric or pseudo-Hermitian.  
We thus instead consider the Hamiltonian shifted by the average-damping envelope for our two modes:
\begin{equation}
    \label{eqn:hShiftedAG}
    \hat{\underline{h}} \equiv \hat{h}_{2 \times 2} - \frac{A_1 \omega_1 + A_2 \omega_2}{2},
\end{equation}
where $A_{\imath} \equiv \langle \phi_{\imath}| \hat{A} | \psi_{\imath} \rangle$. Specializing to the case of the hybridization of two decoupled precessional modes, this shifted Hamiltonian may be calculated explicitly, giving us
\begin{equation}
    \label{eqn:hShiftedAG2}
    \hat{\underline{h}} = \hat{\undertilde{h}} + \frac{A \omega_1 - A \omega_2}{2} \sigma_z,
\end{equation}
where $\hat{\undertilde{h}}$, the projected Hamiltonian in the absence of local damping, is pseudo-Hermitian. $\sigma_z$ is the $z$ Pauli matrix.  
When Eq.~(\ref{eqn:crossingcondition}) holds, $ \omega_2 - \omega_1$ will be of the same order as $\alpha'$, and we have
\begin{equation}
    \label{eqn:A12Condition}
    A_1 \omega_1 - A_2 \omega_2 = (A_1 - A_2) \omega + O(\alpha'^2).
\end{equation}
Thus, the dynamics are pseudo-Hermitian to first order in coupling $\alpha'$, if the local dampings $A_\imath$ are identical.

As a final example, we consider colinear magnetic fields with local Gilbert damping $\alpha_\imath$ and coupling $\hat{G}^-$ given by Eq.~(\ref{eqn:exampleDefa}).  The level splitting of this system is plotted in Fig.~\ref{fig:gilPTbreak}, with the blue curve showing the case of symmetric damping and the green showing asymmetric damping.  In the former case, we see level splitting at second order in the dynamic parameters.  In the the latter case, despite not having a $\mathcal{PT}$-symmetric spectrum even at the leading order, we still retain some qualitative attractive character near the level crossing, as the real frequencies ``pull" together and the imaginary parts ``bulge" outward (relative to the average damping). 

\section{Discussion}

We have constructed a formalism for determining the spectral hybridization properties of classical spin systems. 
Linearized dynamics exhibit level attraction both when two spins are coupled dissipatively or when the spins are prepared in the nonequilibirum regime. 
Interestingly, pseudo-Hermiticity in the nonequilibrium regime, for nondissipative dynamic interactions, survives the inclusion of any static coupling in the free energy.
In contrast, static couplings, such as the exchange interaction, break Pseudo-Hermiticity for dissipatively coupled spins.

Our results are relevant to recent experiments~\cite{subedi2023magnon, yabinUnpublished} that have observed level repulsion as a result of dynamical interactions at equilibrium, as well as older studies of level attraction~\cite{tserkovnyak2005nonlocal, heinrich2003dynamic}.
Such experiments may also explore the non-equilibrium regime, which could reveal qualitatively different spectral properties, as summarized in Table~\ref{table:hybrid}.

Future work may also find interesting to explore the nonlinear regime.  When local damping is compensated, spontaneous $\mathcal{PT}$ symmetry breaking produces dynamical instabilities that can rearrange the magnetic order and may offer an alternative tool for controlling magnetic switching.  Such novel magnetic switching could be useful for developing neuromorphic~\cite{mohseni2022ising} computing platforms.

Another direction is to consider the many-body limit of dynamically coupled spins.
Our results in the nondissipative case appear robust for arbitrarily large systems and any free energy. However, the pseudo-hermiticity of dissipatively coupled systems appears to be significantly more fragile.
For instance, one could explore the dependence of these pseudo-Hermitian features on the particular lattice structure of a large spin system.  
An exciting direction would be to identify what topological classes~\cite{PhysRevX.9.041015, ding2022non} for non-Hermitian dynamics can be accessed via spin systems.


\acknowledgements

This work was supported by the U.S. Department of Energy, Office of Basic Energy Sciences under Grant No. DE-SC0012190.

\appendix

\section{Onsager reciprocity}

We compute the Onsager matrix in the absence of all dynamical effects and demonstrate Onsager reciprocity.
In the second subsection, we discuss Onsager reciprocity for dynamic torques and show it to be satisfied for Eqs.~(\ref{eqn:exampleDefa}) and (\ref{eqn:exampleDefb}).
In the absence of dynamical effects, Eq.~(\ref{eqn:EoM}) takes the form
\begin{equation}
\label{eqn:EoMNoDynamical}
\dot{\textbf{m}} = -\textbf{B}^{\text{eff}} \times  \textbf{m},
\end{equation}
where we have suppressed the index $\imath$ for simplicity.  For a given spin direction $\textbf{m}$, only the component transverse to the spin direction $\textbf{B}^{\text{eff}}_{ \perp} = \textbf{m} \times \textbf{B}^{\text{eff}} \times \textbf{m}$ contributes to the torque.
This component is linear in the spin fluctuation, allowing us to write
\begin{equation*}
\begin{aligned}
\textbf{B}^{\text{eff}}_{ \perp} \times  \textbf{m} & = \textbf{B}^{\text{eff}}_{\perp} \times  \textbf{m}_{0} + O(\delta \textbf{m}^2) \\
& =  \left[\textbf{B}^{\text{eff}} - \left( \textbf{m}_0 \cdot  \textbf{B}^{\text{eff}}\right)  \delta \textbf{m}_{0}\right] \times \textbf{m}_{0}.
\end{aligned}
\end{equation*}
Here and henceforth, we work to linear order in $\delta \textbf{m}$. The second equality follows from inserting the definition of $\textbf{B}^{\text{eff}}_{ \perp}$ and $\textbf{m} = \textbf{m}_{0} + \delta \textbf{m}$, where $\textbf{m}_{0}$ labels the fixed point. Inserting the definition of $\textbf{B}^{\text{eff}}$ for a free energy $F(\text{m}_0, \delta \textbf{m})$ yields
\begin{equation}
\label{eqn:torqueLinearized}
\begin{aligned}
& =  \frac{1}{S}\left[\frac{\delta F}{\delta(\delta \textbf{m})} - \frac{\delta F}{\delta \text{m}_{0}} \delta \textbf{m} \right] \times \textbf{m}_{0}\\
& = \frac{1}{S}\left[\frac{\delta F}{\delta(\delta \textbf{m})} + \frac{\delta F}{\delta \text{m}_{0}} \frac{\delta \text{m}_{0}}{\delta (\delta \textbf{m})} \right] \times \textbf{m}_{0}\\
& = -S^{-1}\hat{M} \hat{\beta} \delta \textbf{m}.
\end{aligned}
\end{equation}
The second line follows from the constraint $\textbf{m}_{0}^2 + \delta \textbf{m}^2 = 1$, and the third line follows from identifying $\hat{\beta}\delta \textbf{m}$ as the derivative of the free energy subject to this constraint. $\hat{M}$ represents the linear operator $\textbf{m}_{0} \times$, with components given by
\begin{equation}
    \label{eqn:Mdef}
    M_{ab} = -\epsilon_{abc} \text{m}_{0,c},
\end{equation}
which is antisymmetric and odd under time reversal.  Comparing Eqs.~(\ref{eqn:torqueLinearized}) and (\ref{eqn:Gammadef}), we identify the Onsager matrix for the two-spin case in block-diagonal form as 
\begin{equation}
\label{eqn:Gamma0Explicit}
\hat{\Gamma}_0 =\left(\begin{array}{ll}
\hat{M}_1 / S_1 & \ \ \ \ 0\\
\ \ \ \  0 &  \hat{M}_2 / S_2
\end{array}\right),
\end{equation}
which is antisymmetric, thus satisfying Onsager reciprocity.

\subsection*{Onsager reciprocity of $\hat{G}^{\pm}$}

In the presence of a dynamic interaction, the Onsager matrix is modified as in Eq.~(\ref{eqn:GammaFromEoM}), with the reciprocal relations taking the form
\begin{equation}
\label{eqn:OSRWithG}
    \left[(1-\hat{G})^{-1} \hat{\Gamma}_0\right]^{\#\intercal} = (1-\hat{G})^{-1} \hat{\Gamma}_0,
\end{equation}
where the superscript $^{\#}$ indicates time reversal.
By clearing the denominators of Eq.~(\ref{eqn:OSRWithG}) and using $\hat{\Gamma}_0^{\# \intercal} = \hat{\Gamma}_0$, we arrive at

\begin{equation}
\label{eqn:OSREquivalent}
    \left(\hat{G} \hat{\Gamma}_0\right)^{\#\intercal} = \hat{G} \hat{\Gamma}_0.
\end{equation}

\noindent Thus, subject to $\hat{\Gamma}_0$ being Onsager reciprocal, proving the Onsager reciprocity of $\hat{G}$ reduces to proving Eq.~(\ref{eqn:OSREquivalent}).
From this, it is clear that any linear combination of Onsager-reciprocal dynamic interactions will also be Onsager reciprocal. 

For the interaction given by Eq.~(\ref{eqn:exampleDefa}), the matrix in Eq.~(\ref{eqn:OSREquivalent}) becomes
\begin{equation}
    \label{eqn:GGammaDissipative}
    \hat{G} \hat{\Gamma}_0 = -\frac{\alpha'}{S} \left(\begin{array}{ll}
 \ \ \ 0 &  \hat{M}_1 \hat{M}_2 \\
\hat{M}_2 \hat{M}_1 &  \ \  \ 0 \\
\end{array}\right).
\end{equation}
Because $\hat{M}_{\imath}$ is both antisymmetric and odd under time reversal, it follows that $\hat{G}\hat{\Gamma}_0$ is both symmetric and even under time reversal, thus satisfying Onsager reciprocity.  

For the interaction given by Eq.~(\ref{eqn:exampleDefb}), it is sufficient to check Onsager reciprocity for the case $\eta_2 = 0$, while the proof for the case that $\eta_1 =0$ is identical. 
In this case, we have
\begin{equation}
    \label{eqn:GGammaNonDissipative}
    \hat{G} \hat{\Gamma}_0 = \frac{\eta_1}{S} \left(\begin{array}{ll}
 \ \ \ \ \  0 &  \hat{M}_1 \hat{M}_1 \hat{M}_2 \\
\hat{M}_2 \hat{M}_1 \hat{M}_1 &  \ \ \ \  \ 0 \\
\end{array}\right).
\end{equation}
Once again, the factors $\hat{M}_{\imath}$ are antisymmetric and odd under time reversal, from which it follows that Eq.~(\ref{eqn:OSREquivalent}) is satisfied. 

\section{Overview of pseudo-Hermiticity}

Pseudo-Hermitian Hamiltonians represent a class of non-Hermitian matrices satisfying Eq.~\eqref{eqn:pseudoHermitian} with a pseudometric $\hat{\Theta}$ \cite{mostafazadeh2008pseudo}. 
Taking the Hermitian conjugate of Eq.~(\ref{eqn:pseudoHermitian}) allows us to see that if $\hat{\Theta}$ is a good pseudometric for $\hat{h}$ then so is $\hat{\Theta}^{\dagger}$.  
Furthermore, taking a linear combination of the two generally allows us to construct a pseudometric that is explicitly Hermitian and thus has real eigenvalues.
$\hat{h}$ in general need not have real eigenvalues or even be diagonalizable.  
Points where it is not diagonalizable are called exceptional points.

An important special case occurs when a Hermitian $\hat{\Theta}$ has definite sign, usually chosen to be positive; these Hamiltonians are called quasi-Hermitian, and and may be rescaled as follows:
\begin{equation}
    \label{eqn:Hrescaled}
    \hat{H} = \hat{\Theta}^{1/2} \hat{h} \hat{\Theta}^{-1/2},
\end{equation}
where $\hat{\Theta}^{1/2}$ is Hermitian.  
The Hamiltonian $\hat{H}$ is Hermitian, and is therefore guaranteed to be diagonalizable and possess real eigenvalues.
For a summary of the distinctions between pseudo-Hermitian, quasi-Hermitian, and Hermitian, see Table \ref{table:1}.

In general, a pseudo-Hermitian Hamiltonian implies the existence of a conserved quantity defined by
\begin{equation}
    \label{eqn:pseudoNorm}
    ||\psi ||  = \langle \psi | \hat{\Theta} | \psi \rangle,
\end{equation}
which is analagous to probability in Hermitian quantum mechanics.  
As we show later in this Appendix, away from any exceptional points, if $| \psi \rangle$ is an eigenstate of $\hat{h}$ with eigenvalue $\omega$, then Eq.~(\ref{eqn:pseudoNorm}) will be zero iff $\omega$ has nonzero imaginary part.  
Thus, any state with real $\hat{h}$ eigenvalue may be designated as positive or negative according to the sign of $||\psi||$.

Pseudo-Hermiticity also implies the existence of a $\mathcal{PT}$ symmetry. Specifically, we have
\begin{equation}
    \label{eqn:Hstar}
    \mathcal{PT}\hat{h}(\mathcal{PT})^{-1} = \hat{h},
\end{equation}
where $\mathcal{PT} = \hat{R} \mathcal{K} \hat{\Theta} $, $\hat{R}$ is a similarity transformation satisfying $\hat{R}\hat{h}^{\intercal}\hat{R}^{-1} = \hat{h}$, and $\mathcal{K}$ is the antilinear operator representing complex conjugation.  
$\mathcal{PT}$ symmetries, originally studied for their application to open quantum systems {\cite{bender2019pt}}, are now commonly applied to optics.  
If a Hamiltonian possessing such a symmetry has an eigenvalue $\omega$ then it must possess the eigenvalue $\omega^*$ as well, and it is said to have a spontaneously-broken $\mathcal{PT}$ symmetry when $\Im \omega \neq 0$ for at least one eigenvalue.  
As summarized in Table \ref{table:1}, $\mathcal{PT}$-symmetry breaking does not occur for quasi-Hermitian Hamiltonians. 
As discussed in Ref.~\cite{bender2019pt}, in the $\mathcal{PT}$-unbroken phase, it is possible to define a $\mathcal{CPT}$ inner product, under which the dynamics become Hermitian.

\subsection*{Riesz basis}

Following Refs.~\cite{mostafazadeh2008pseudo, mostafazadeh2002pseudo}, we define the Riesz basis and demonstrate the connection between $\mathcal{PT}$-symmetry breaking and the zeros of the conserved quantity $\left| \left| \psi \right| \right|$. 
The pseudo-Hermitian condition given in Eq.~(\ref{eqn:pseudoHermitian}) is reminiscent of a symmetry of the Hamiltonian: $\hat{S} \hat{h}\hat{S}^{-1} = \hat{h}$.
While the latter implies the existence of a basis that simultaneously diagonalizes $\hat{h}$ and $\hat{S}$, the former implies a weaker condition: the existence of a specific Riesz basis.  

As discussed in Ref.~\cite{mostafazadeh2008pseudo}, the Riesz basis consists of two complete sets of vectors, $\{|\psi_n\rangle, |\tilde{\psi}_{\nu}\rangle \}$ and $\{|\phi_n \rangle, |\tilde{\phi}_{\nu}\rangle \}$, where $n = 1,...,N$ and $\nu = \pm 1, ... , \pm \tilde{N}$, with $N + \tilde{N}$ equaling the dimension of the vector space.  
Each set individually forms a basis for the vector space, while the inner products between elements of the two sets satisfy

\begin{equation}
    \label{eqn:innerPapp}
    \begin{aligned}
 \langle\psi_n {\mid} \phi_m\rangle & =\delta_{n m} ,\,\,\,
\langle\tilde{\psi}_\nu {\mid} \tilde{\phi}_\mu\rangle =\delta_{\nu \mu}, \\
\langle\psi_n {\mid} \tilde{\phi}_\nu \rangle & =\langle\tilde{\psi}_\nu {\mid} \phi_n\rangle=0.
\end{aligned}
\end{equation}

\noindent Notably, Eq.~(\ref{eqn:innerPapp}) leaves open the choice for the normalization of either one of these bases but not both.
According to Ref.~\cite{mostafazadeh2002pseudo}, for any pseudo-Hermitian Hamiltonian away from an exceptional point, there exists a Riesz basis such that 
\begin{equation}
    \label{eqn:hriesz}
    \begin{aligned}
 \hat{h}  = & \sum_{n=1}^{N} \omega_{n}\left|\psi_{n}\rangle\langle\phi_{n}\right| \\ 
& +\sum_{\nu=1}^{\tilde{N}}\left(\tilde{\omega}_\nu|\tilde{\psi}_{\nu}\rangle\langle\tilde{\phi}_{\nu}|+\tilde{\omega}_\nu^*|\tilde{\psi}_{-\nu}\rangle\langle\tilde{\phi}_{-\nu}|\right), \\
 \hat{\Theta}  = & \sum_{n=1}^{N} \theta_{n}|\phi_{n}\rangle\langle\phi_{n}| \\ &+\sum_{\nu=1}^{\tilde{N}}\left(|\tilde{\phi}_{\nu}\rangle\langle\tilde{\phi}_{-\nu}|+|\tilde{\phi}_{-\nu}\rangle\langle\tilde{\phi}_{\nu}|\right),
\end{aligned}
\end{equation}
where $\omega_n$ and $\tilde{\omega}_{\nu}$ represent the real and complex eigenfrequencies (if any), respectively. $\theta_n$ is the $\hat{\Theta}$ expectation value for $|\psi_n\rangle$, with the normalization of $|\psi_n\rangle$ chosen such that $\theta_n = \pm 1$.
We can see by direct calculation that  $\{|\psi_n\rangle, |\tilde{\psi}_{\nu}\rangle \}$ and $\{|\phi_n\rangle, |\tilde{\phi}_{\nu}\rangle \}$ are the right and left eigenvectors of $\hat{h}$, respectively.
Furthermore, these sets are related by 
\begin{equation}
\label{eqn:LeftfromRightApp}
    \hat{\Theta} |\psi_n\rangle = \theta_n |\phi_n\rangle, ~~~  \hat{\Theta}|\tilde{\psi}_{\nu}\rangle = |\tilde{\phi}_{-\nu}\rangle.
\end{equation}
From Eq.~(\ref{eqn:hriesz}), we can also show by explicit calculation that the conserved quantity defined in Eq.~(\ref{eqn:pseudoNorm}) is given for the right eigenvectors by $\left| \left| \psi_n \right| \right| = \theta_n$ and $| | \tilde{\psi}_{\nu} | | = 0$. 
Thus, $\left| \left| \psi \right| \right|$ is zero only for eigenmodes with imaginary frequencies.

\section{Explicit calculation of examples}

We here compute explicitly the matrices relevant to the examples discussed in Sec.~V.
The unperturbed part of the Hamiltonian is given explicitly by 
\begin{equation}
\label{eqn:h0explicit}
\hat{h}_0 = \omega_1 P_1 \otimes \sigma_y +  \omega_2 P_2 \otimes \sigma_y,
\end{equation}
where the first tensor factor corresponds to the two macrospins while the second corresponds to the Cartesian fluctuation components. 
$P_{\imath} = [1 + (-1)^{\imath-1}\sigma_z]/2$ is the projector onto the space of the fluctuations of spin $\imath $. 
The free-energy curvature matrix $\hat{\beta}$, defined in Eq.~(\ref{eqn:betadefF}), is given by 
\begin{equation}
\label{eqn:betaexplicit}
\hat{\beta} = \zeta_1\omega_1 S_1 P_1 \otimes 1 +  \zeta_2 \omega_2 S_2 P_2 \otimes 1.
\end{equation}
The linearization of Eq.~(\ref{eqn:exampleDefb}) leads to the interaction part of the Hamiltonian given by
\begin{equation}
\label{eqn:NonDissipativeGExplicit}
\begin{aligned}
    \hat{G}\hat{h}_0 = & \frac{\eta_1 \zeta_1 + \eta_2 \zeta_2 }{S} (\omega_1 S_1 \zeta_2 \sigma_- + \omega_2 S_2 \zeta_1 \sigma_+) \otimes \sigma_y.
\end{aligned}
\end{equation}
The calculation for the dissiptive interaction given by Eq.~(\ref{eqn:exampleDefa}) is similar, with the interaction part of the Hamiltonian now given by
\begin{equation}
\label{eqn:DissipativeGExplicit}
\hat{G}\hat{h}_0 = -\frac{i\alpha'}{S} (\zeta_1 \omega_2 S_2 \sigma_+ + \zeta_2 \omega_1 S_1 \sigma_-) \otimes 1.
\end{equation}

\nocite{*}
\bibliography{aapmsamp}

\end{document}